\documentstyle[preprint,aps]{revtex}
\begin{document}
\draft
\title{\bf A First-Landau-Level Laughlin/Jain Wave Function
for the Fractional Quantum Hall Effect}
\author{Joseph N. Ginocchio}
\address{T-5, Los Alamos National Laboratory \\
\it Los Alamos, New Mexico  87545}
\author{W.C. Haxton}
\address{Institute for Nuclear Theory, Box 351550 \\
and Department of Physics, Box 351560 \\
University of Washington, Seattle, Washington  98195-1550}
\date{\today}
\maketitle
\begin{abstract}
We show that the introduction of a more general closed-shell operator
allows one to extend Laughlin's wave function to account
for the richer hierarchies (1/3,
2/5, 3/7 $\dots$\,; ~ 1/5, 2/9, 3/13, $\dots$\,, etc.) found
experimentally.  The construction identifies the
special hierarchy states with condensates of correlated electron clusters.
This clustering implies a single-particle algebra within the first Landau
level (LL) identical to that of multiply filled LLs in the integer quantum Hall effect.
The end result is a simple generalized wave function that reproduces
the results of both Laughlin and Jain, without reference to higher
LLs or projection.
\end{abstract}
\pagebreak

One of the more intriguing problems in condensed matter physics has been the
explanation of the fractional quantum Hall effect (FQHE) [1,2] and the
connection of this phenomenon with the integer case (IQHE) [1,3].  While
the IQHE
can be understood [4] qualitatively as a noninteracting electron problem,
the FQHE must be connected
with breaking of the degeneracy of the noninteracting ground state by the
electron-electron Coulomb interaction.  
Laughlin provided a simple and physically appealing ground-state wave function for
fractional fillings of the form $1/m$, $m$ an odd integer [5].  Others have
attempted to extend his ansatz in order to account for the more complex
pattern of minima in the resistivity seen experimentally, particularly the
strong features at fractional fillings 1/3, 2/5, 3/7,...[6-9].
In particular, one effort that is remarkable for its numerical success is that of Jain and his collaborators [8], who
introduced excitations into successively higher LLs that were later
eliminated by numerical projection.  Yet the need for
such excitations is somewhat puzzling:
they play no role in the final answer, nor is there a natural argument
associating the large magnetic gaps of the IQHE with the small
Coulomb splittings within the first LL.

In this letter we construct a physically appealing generalization of
Laughlin's ansatz that we believe greatly helps in reconciling the work 
of Laughlin, Jain, and others.  The construction is entirely in the first LL
and extends Laughlin's wave function in a remarkably simple way:
a closed $\ell$-shell in his treatment is replaced by a closed $\ell s$-shell,
where 2$s$+1 counts the number of electrons involved in the underlying
clusters (or composites).  When such wave functions are recast in their
corresponding $(ls)j$ form, a shell structure analogous to filled LLs in the IQHE
emerges, but with shell gaps determined by a pairing energy associated
with the Coulomb force, and not by the magnetic energy.  The net
result is a simple generalization of Laughlin's wave function that
also reproduces the results of Jain without reference to higher LLs or
projection.

We follow Haldane [6] in confining the electrons to the surface of a sphere,
where they move under the influence of a perpendicular magnetic field generated
by a monopole at the sphere's center.  The Dirac monopole quantization
condition requires $\phi = 2S\phi_0$ where $\phi_0 = hc/e$ is the elementary
unit of magnetic flux, with $2S$ an integer.  The single-particle wave
functions are Wigner D-functions
\begin{equation}
D^{(L)}_{S,q}(\phi,\theta,0)
\end{equation}
where $L$ is the Landau level index, $L=S,\;S+1,...$.  Thus there are $2S+1$
degenerate single-particle wave functions in the first LL,
$2S+3$ in the second, etc.  The wave functions for the first LL
can be written as a monomial of power $2S$ in the elementary spinors $u_q$
\begin{equation}
D^{(S)}_{S,q} = \left[{(2S)!\over (S+q)!(S-q)!}\right]^{1/2} u^{S+q}_
{1/2} u^{S-q}_{-1/2}\, \equiv \,u^{2S}_q
\end{equation}
where
\begin{equation}
u_q(\phi,\theta) = D^{(1/2)}_{1/2,q} (\phi,\theta) =
\left\{\vcenter{\halign{ $ # $ \hfill\qquad &
$ # $ \hfil \cr
\cos(\theta/2)
e^{i\phi/2},& q=1/2 \cr
\sin(\theta/2)\;e^{-i\phi/2}, & q=-1/2 \cr}}\right.  \eqnum{3}
\end{equation}
(identical to the $u$ and $v$ of Haldane [6]).  The Laughlin wave
function [5,6] for $N$ electrons corresponds to $m$ powers of the closed
shell, $m$
odd,
\begin{equation}
L_m(N) = \left[\prod^N_{i<j} u(i)\cdot u(j)\right]^m \equiv
\left[\prod^N_{i<j} u(i)\cdot u(j)\right] L^{sym}_{m-1}(N)\eqnum{4a}
\end{equation}
where the factoring of this expression into a closed-shell operator
acting on a symmetric wave function will prove useful later.
Here $u(i) \cdot u(j)$ = $u_{+1/2}(i) u_{-1/2}(j) - u_{-1/2}(i) u_{+1/2}(j)$ is the
usual scalar product of two spin-1/2 tensors.  Hence Eq. (4a) has 
total angular momentum zero and is the analog
of a translationally invariant state in the plane.  As there are $m(N-1)$
appearance of each spinor $u(i)$
\begin{equation}
2S = m (N-1). \eqnum{5a}
\end{equation}
This wave function is identified with fractional filling $1/m$:
$N/(2S+1) \to 1/m$ for large $N$.
Because it contains $m$ powers of $u(i) \cdot u(j)$, two-particle angular
momenta are restricted to $J_{ij} \leq 2S - m$.  This excludes
contributions from the largest Coulomb matrix elements $\langle (SS)
J_{ij}|V_c|(SS) J_{ij}\rangle$ corresponding to short-range electron-electron
interactions, thus helping to minimize the energy.  The wave function
is ``incompressible"
because any reduction in $S$, for fixed $N$, forces electrons into
configurations with $J_{ij}
> 2S - m$.

Below we will consider the process of compressing a Laughlin state, labeled by
$N$ and $m$, by successively reducing the magnetic flux (and thus $S$),
ultimately reaching the next Laughlin state $(N,m-2)$.  The task before
us is to identify other incompressible states encountered in this
process - nondegenerate ground states distinguished energetically by
their special symmetry.  We begin by describing a classical caricature
of this problem that may help the reader conceptually.  Envision
charges living on a one-dimensional regular lattice.  If we start with
6 charges and 11 lattice sites, the obvious minimum energy configuration
has the charges spaced uniformly, on sites 1,3, ...,11.  In this configuration
no two charges appear on
neighboring sites, a condition analogous to the two-body Laughlin restriction
on $J_{ij} > 2S-m$.  Now remove one lattice site at a time
and look for similar unique configurations of lowest energy.  The first
such case arises for 8 sites, with the charges clustered in pairs,
occupying sites 1,2,4,5,7,8.  This configuration is completely specified
by a condition on three-body correlations, that no three charges are
allowed to occupy contiguous sites.  For 7 sites another such state
is found, comprised of clusters of three: the occupied sites are
1,2,3,5,6,7, and
no four particles occupy continguous sites.
The pattern of denser states, clustering, and more complicated many-particle
correlations is clear.

Now consider the analogous progression through states of increasing
density in the FQHE, e.g., the 1/3, 2/5, 3/7, ... hierarchy.
Eq. (4a) states that the first term in this series is produced by
the action of an antisymmetric closed-shell operator acting
on $L^{sym}_2$, the (symmetric) half-filled shell to which the
series converges.  This operator produces $N-1$ units of magnetic flux.
This suggests constructing analogous
antisymmetric operators for the ``compressed" states 2/5, 3/7,... corresponding to reduced magnetic flux.
A scalar operator that destroys magnetic flux is readily constructed: defining 
$d_q = (-1)^{1/2 + q} {d \over du_{-q}}$, one finds $d(i) \cdot d(j) \, u(i)
\cdot u(j) =2.$  This operator can be applied in such a way that it reduces
the magnetic flux by one unit, provided $N$ is even
\begin{equation}
\left[d(1) \cdot d(2)\, \dots\, d(N-1) \cdot d(N) \right]
\, L^{sym}_2 \eqnum{6}
\end{equation}
  
The derivatives produce a condensate of particle pairs that - by necessity due to the reduced magnetic
flux - are unfavorably correlated spatially:\, $d(i) \cdot d(j)$ acting on
$(u(i)^{N-1} \otimes u(j)^{N-1})_{J_{ij}}$ destroys one power of each spinor,
reducing $S$, but does not change $J_{ij}$.
This is reminiscent of our classical configuration of 6 clustered charges on 8
lattice sites, with the important difference that we have not yet imposed
any condition that will keep the clusters separated.  Such separation
is important in minimizing the energy.  Now Laughlin's
closed-shell operator separates single electrons
\begin{equation}
\prod^N_{i < j = 1} u(i) \cdot u(j) = {\cal A}\left[\prod^N_{i < j = 1} \, u_{-1/2} (i) u_{+1/2} (j) \right]
\eqnum{7a}
\end{equation}
where the antisymmetrization operator ${\cal A}$ has been introduced
to make the corresponding generation for pairs obvious.  An operator that
produces and separates two-particle clusters is thus
\begin{equation}
{\cal A} \left[\prod^{N/N_c}_{I<J} U_- (I) \, U_+
(J) \,
\left(d(1) \cdot
d (2) \dots d(N-1) \cdot d (N) \right) \right] 
\eqnum{7b}
\end{equation}
where $N_c$=2 and $U_-(I=1) = u_{-1/2} (1) \, u_{-1/2} (2), \, U_-(I=2) = u_{-1/2} (3) \,
 u_{-1/2}
(4)$, etc.  Note that $U_-(I) \, U_+(J)$, acting on a four-particle
m-scheme configuration, increases 2$S$ by one unit, lowers the
magnetic quantum numbers of the particles in cluster $I$, and
raises those in cluster $J$.  Thus it spreads the clusters in
azimuthal space.  The generalization for all pairs in Eq. (7b)
displaces each of the clusters relative to one another and
increases 2$S$ by $N/N_c-1$.  Thus this operator is responsible
for a fractional change in $2S+1$ of $1/N_c$ relative to that
produced by Laughlin's closed-shell operator of Eq. (7a).

We introduced an additional quantum number in Eq. (7b), $N_c$, the clustering
size.  Clearly, for all $N_c$ for which $N/N_c$ is an integer, we can repeat
the arguments given above.  The generalization of the derivative
pairs $d(1) \cdot d(2)$ to any clustering size is $L^1_d (I) \equiv
\prod^{N_c}_{i < j = 1} d(i) \cdot d(j)$, where $I$ represents the set of
particles $\{1, \dots, N_c\}$.  This provides one factor of $d(i) \cdot d(j)$
for each possible pairing of particles in the cluster.
The corresponding generalization of $U_-(I)$ is
clearly $u_{-1/2} (1) \dots u_{-1/2} (N_c)$.
Thus for arbitrary $N_c$ we obtain
\begin{equation}
{\cal A} \left[
\prod^{N/N_c}_{I < J} U_-(I) U_+(J) \prod^{N/N_c}_{I=1} L_d^1 (I) \right]
\eqnum{7c}
\end{equation}
which reduces to Eqs. (7a) and (7b) for $N_c$ = 1 and 2, respectively.
The substitution of Eq. (7c) for (7a) in the right-hand side of Eq. (4a)
gives our generalized Laughlin wave function.
  
Eq. (7c), deduced from rather straightforward physical arguments, can be
rewritten in a form that better illustrates its simple connections to
Laughlin's wave function.  We introduce an angular momentum $(\ell,m_i)$ for the $i$th electron,
where 2$\ell$+1 = $N/N_c$ is the number of clusters, and a spin $(s,q_i)$ with
2$s$+1 = $N_c$, the number of electrons in each cluster.  Thus the total number of distinct
pairs of magnetic indices $(m,q)$ is $(2 \ell+1)(2s+1) = N$.  Then
Eq. (7c) can be rewritten and substituted into Eq. (4a) to give
\begin{equation}
L_{m,N_c}(N) = \left[ \sum_{m's,q's} \epsilon_{M_1 \, \ldots \, M_N} \,
u^{2 \ell}_{m_1} (1) \, \ldots \, u^{2 \ell}_{m_N} (N) \, 
d^{2s}_{q_1} (1) \, \ldots \, d^{2s}_{q_N} (N) \right] L^{sym}_{m-1}
\eqnum{4b}
\end{equation}
where $\epsilon$ is the antisymmetric tensor with $N$ indices and $M_i = (m_i,q_i)$.  Thus
Eq. (4b) is obtained from from Laughlin's wave function (Eq. (4a)) by replacing
a closed $\ell$ shell ($2 \ell+1=N$) by a closed
$\ell s$ shell ($(2 \ell+1)(2s+1)=N$), where the spin is associated with the
destruction of $2s$ units of magnetic flux accompanied by   
the clustering of sets of 2$s$+1 electrons.
Like Eq. (4a), it has total angular momentum zero.  This $\ell s$
operator, written as a Slater determinant, was first
introduced in Ref. [10] from arguments based on symmetry.  
In deriving Eq. (7c) we have shown it is the mathematical manifestation
of the underlying clustering.

As the derivatives decrease the number of flux quanta,
the corresponding generalization of Eq. (5a) is
\begin{equation}
2S = (m-1)(N-1) + 2 \ell - 2s = (m-1)(N-1) + {N \over N_c} - N_c \eqnum{5b}
\end{equation}
where $m$ is an odd integer and $N_c$ can take on any integral value
that divides evenly into $N$. 
For fixed cluster size $N_c$ and large $N$,
the fractional filling becomes $N/(2S+1) \to N_c/((m-1)N_c+1)$.  Thus
the $m$=3 hierarchy corresponds to fractional fillings 1/3, 2/5, 3/7,
$\ldots \,$; $m$=5 gives 1/5, 2/9, 3/13, $\ldots \,$; etc.
These series converge to $1/(m-1)$ = 1/2,1/4, etc., from the low-density
side.  One can also consider large $N$ for a fixed number of clusters $\bar N_c = N/N_c$, $N/(2S+1) \to
\bar N_c/((m-1) \bar N_c-1)$.  This yields the $m$=3 series converging
to the half-filled shell from the high-density side, 1, 2/3, 3/5, $\ldots \,$;
the $m$=5 series 1/3, 2/7, 3/11, $\ldots \,$; etc.
For finite $N$, the natural division between the low- and high-density
series is $N=N^2_c$ or, equivalently, $\ell=s$.

Thus we see the $m$ series, generated by antisymmetric closed-shell
operators of Eq. (7c) acting on the $1/(m-1)$-filled shell, span the
fractional fillings between 1/$m$ and $1/(m-2)$.
If one steps through this evolution for
fixed N by gradually reducing the magnetic flux,
Eq. (7c) states that the cost of the resulting compression is the condensation of clusters of
electrons that are unfavorably correlated spacially: within each
cluster the allowed two-electron angular momenta $J_{ij}$ are those
of the $m$-2 Laughlin state.  The hierarchy states $(m,N_c)$
correspond to those special geometries where the ground state is
filled by a condensate of clusters, each containing $N_c$ electrons.  Any further compression
of the system forces clusters of $N_c+1$ electrons into unfavorable
correlations characteristic of the denser $m-2$ Laughlin state,
with a corresponding increase in the energy/particle.  In this 
sense the states are incompressible.
Continued compression produces larger clusters, with the final step
being the single-cluster $1/(m-2)$ Laughlin state.

The Laughlin states are eigenstates of a two-body interaction
that is infinitely repulsive for $J_{12} > 2S-m$.  They also
exclude three-body correlations for which $J_{123} > 3S-3m$ and,
in general, $J_{1...n} > nS-mn(n-1)/2$.  
The generalized Laughlin state for a given ($S,N$) is an eigenfunction of 
an $N_c+1$-body interaction that is infinitely repulsive for
$J_{1...N_c+1} > (N_c+1)(S-(m-2)N_c/2-1)$ [11].  This constraint is
more severe than the corresponding constraint for the ($m$-2) Laughlin
state.  Thus, while the $N_c$ electrons within each cluster are forced
into correlations characteristic of the denser $m-2$ Laughlin state,
the system maintains correlations among $N_c$+1 or more particles
more favorable than those of the $m$-2 Laughlin state.  All of this
is strikingly similar to our classical lattice caricature.

Eq. (4b) can be easily evaluated using
\begin{equation}
d^\kappa_q u^n_{q_1} \propto (-1)^{{\kappa \over 2} + q} \,
\left[{\kappa !
\over ({\kappa \over 2} + q)! ({\kappa \over 2} - q)!}\right]^{1/2}
\left[{({n \over 2} + q_1)! ({n \over 2} - q_1)! \over ({n - \kappa \over 2} +
q +q_1)! ({n - \kappa \over 2} - q - q_1)!}\right]^{1/2} \, u^{n-
\kappa}_{q+q_1}. \eqnum{10}
\end{equation}
Table 1 gives the overlaps with exact wave functions obtained
by numerical diagonalization [12].  We include results through $N = 10$ and
$2S = 24$, excluding trivial cases $(N \leq 3$ or $\bar N = 2S + 1 - N \leq
3$).  The overlaps for $N_c \geq 2$ are $\geq$ 0.993 in all cases.
The corresponding results for Jain's projected wave functions on the sphere 
are given in [8] for $(2S,N)$ = (11,6), (19,6), and (16,8),
corresponding to states of filling 2/5, 2/7, and 2/5, respectively.
Our results are identical to four significant digits in the 2/5
cases, and differ by 0.0008 for the 2/7 state.  

The reason for this agreement is that the algebra of 
filled LLs of the IQHE, which Jain employs, is identical to an algebra
that exists entirely within the first LL, imposed by the clustering:
the closed $\ell s$ shell of Eq. (4b) can be immediately recoupled in the form
$(\ell s)j$, forming a set of $N_c=2s+1$ closed $j$ shells with $|\ell - s| \le
j \le \ell + s$, that is, a shell structure identical to $N_c$ filled
LLs of the IQHE, where the first LL contains $2(\ell - s) +1$ electrons.
We have shown that the operations employed by Jain - construction
of higher LL wave functions followed by projection - are mimicking
the effects of electron clustering within the first LL, a phenomenon
we have argued is a natural consequence of the compression of a
Laughlin state.

The origin of shell gaps within the first LL is easy to see.  As
the general case is similar, we illustrate the physics for $N_c$=2.
The operator of Eq. (7c) is then a product of two closed $(\ell s=1/2)j$ shells, the lower one consisting of 
single-particle operators $[u^{2 \ell}(i) \otimes d(i)]_{j = \ell - 1/2} =
u^{2 \ell -1}(i) u(i) \cdot d(i) = u^{2 \ell -1}(i)$.  Here we have
noted that the derivatives act on $L^{sym}_{m-1}(N)$,
for which $u(i) \cdot d(i)$ is the identity operator.  The upper 
shell operators are $[u^{2 \ell}(i) \otimes d(i)]_{j=\ell +1/2} =
[u^{2 \ell -1} \otimes [u(i) \otimes d(i)]_1]_{j=\ell + 1/2}$.  But the
operator $[u(i) \otimes d(i)]_{1}$, acting on $L^{sym}_{m-1}$,
destroys an antisymmetric $u(i) \cdot u(j)$ pair, replacing it with the symmetric product $[u(i) \otimes u(j)]_{1}$,  
which then allows the maximum value of $J_{ij}$ to increase by one.
Thus electrons $i$ and
$j$ can approach more closely, at a corresponding cost in the energy.
The general $N_c$ case is similar: each successive filled $(\ell s)j$
shell involves one additional pair-breaking operator: it is the
energy of the broken pair that generates the shell gap [11].

In summary, we have shown than a particularly simple generalization
of Laughlin's wave function - replacement of a filled $\ell$ shell 
by a filled $\ell s$ shell - is a natural consequence of the
clustering of electrons when a Laughlin state is compressed.  This clustering
implies an $(\ell s)j$ algebra within the first LL level, analogous
to multiply filled LLs in the IQHE, but with
shell energies indexed by the number of broken $u(i) \cdot u(j)$ pairs.
We have thus reproduced the results of both Laughlin and Jain in
a simple first-LL wave function, while providing an appealing
physical argument in support of Jain's construction.

The clustering also offers an interesting perspective on the
spectroscopy of excited states in the FQHE.  For example, Rezayi
and Read [13] accounted for the first excited band of a series
of ``half-filled" systems as quasiparticle-hole valence excitations of
a postulated shell Hamiltonian $H$=$\vec L^2$.  We recognize this 
as the special case of the $(\ell s)j$ algebra where $\ell = s$,
so that $j=L$=0,1,2,... and $N$=1,9,16,...   
All other $m=3$ hierarchy states can be arranged in similar
series converging to the half-filled shell, corresponding to 
increasing $\ell$ with $\ell - s$ held fixed, and the arguments of Ref. [13]
can be applied to each [11].  Thus the shell structure and shell
gaps associated with electron clustering provide the starting 
point for exploring FQHE spectroscopy quite generally.

We thank David Thouless for many helpful discussions, and J. K. Jain
for several comments.  This work was supported in part by the U.S.
Department of Energy.

Email addresses:  gino@t5.lanl.gov,\,
                  haxton@phys.washington.edu
\pagebreak
\begin{table}
\caption{Overlaps [12] $|\langle \Psi_{exact}|\Psi_{m, N_c} \rangle|$ for
the generalized Laughlin wave functions of Eq. (4b).
The third column lists the lesser of $N_c$ and
$\bar N_c$, with the latter appearing in ().  Both corresponding
fractional fillings are provided 
when $N_c$ = $\bar N_c$.
The original Laughlin wave functions are those with $N_c$ = 1.}
\vspace{18pt}
$$\vbox {\tabskip 2em plus 3em minus 1em
\halign to \hsize{\hfil #\hfill && #\hfil \hfil \cr
N&2S&$N_c(\bar N_c)$&$m$&filling&$|\langle \Psi_{ex}|\Psi\rangle|$ \cr
\noalign{\hrule}
\noalign {\vskip 12pt}
4&9&1&3&1/3&.9980 \cr
4&12&2&5&2/9,2/7&.9999 \cr
4&15&1&5&1/5&.9841 \cr
4&18&2&7&2/13,2/11&.9992 \cr
4&21&1&7&1/7&.9741  \cr
5&12&1&3&1/3&.9991  \cr
5&20&1&5&1/5&.9974 \cr
6&9&(2)&3&2/3&.9965 \cr
6&11&2&3&2/5&.9998 \cr
6&15&1&3&1/3&.9965 \cr
6&19&(2)&5&2/7&.9964\cr
6&21&2&5&2/9&.9928 \cr
7&18&1&3&1/3&.9964 \cr
8&12&(2)&3&2/3&.9982 \cr
8&16&2&3&2/5&.9996 \cr
8&21&1&3&1/3&.9954 \cr
9&16&3&3&3/7,3/5&.9994 \cr
9&24&1&3&1/3&.9941 \cr
10&15&(2)&3&2/3&.9940 \cr
10&21&2&3&2/5&.9980 \cr
\noalign {\vskip 12pt}
}}$$

\end{table}
\pagebreak

\end{document}